\documentclass[twocolumn,twoside,prl,aps,superscriptaddress,showpacs]{revtex4}
\usepackage{amsmath,mathrsfs,amsbsy,color,graphicx,bm,amsthm,amsfonts}
\usepackage{units}
\usepackage{bbm}

\begin{document}

\title{All-optical routing of single photons with multiple input and output
ports by interferences}
\author{Wei-Bin Yan}
\email{yanweibin1983@dlnu.edu.cn}
\affiliation{Liaoning Key Lab of Optoelectronic Films \& Materials, School of Physics and
Materials Engineering, Dalian Nationalities University, Dalian, 116600, China}
\author{Bao Liu}
\affiliation{Beijing Computational Science Research Center (CSRC), Beijing 100084, China}
\author{Ling Zhou}
\email{zhlhxn@dlut.edu.cn}
\affiliation{School of Physics and Optoelectronic Technology, Dalian University of
Technology, Dalian, 116024, China}
\author{Heng Fan}
\email{hfan@iphy.ac.cn}
\affiliation{Beijing National Laboratory for Condensed Matter Physics, Institute of
Physics, Chinese Academy of Sciences, Beijing 100190, China}

\begin{abstract}
We propose a waveguide-cavity coupled system to achieve the routing of
photons by the phases of other photons. Our router has four input ports and
four output ports. The transport of the coherent-state photons injected through any input
port can be controlled by the phases of the coherent-state photons injected
through other input ports. This control can be achieved when the mean
numbers of the routed and control photons are small enough and require no
additional control fields. Therefore, the all-optical routing of photons can
be achieved at the single-photon level.
\end{abstract}

\maketitle

Quantum network \cite{network} plays an essential role in quantum
information and quantum computation \cite{ic}. The routing capability of
information is a requisite in quantum network. Photons are considered as the
ideal carrier of quantum information. Therefore, the investigation of
all-optical routing of photons at the single-photon level will have direct
application to realize quantum networks for optical quantum information and
quantum computation. Recently, a scheme to achieve all-optical routing of
single photons with two input ports and two output ports has been
demonstrated \cite{dayang}. In their scheme, the control single photons and
routed single photons are connected by an intermediate three-level atom. By
coupling two different atomic transitions, respectively, to the routed and
control photons, the routed single photons can be controlled through
injecting the control single photons. Currently, it will be of interest to
realize the all-optical routing of photons at the single-photon level in
other physical mechanism. Moreover, the all-optical routing with more than
two input and output ports, which is essential for the quantum network,
still needs to be explored.

For these purposes, we propose a scheme to study the all-optical routing of
coherent-state photons with four input ports and four output ports by other
coherent-state photons. It is significant that the all-optical routing of
photons is realized by the interferences depending on the phase differences
between the routed and the control photons. Our scheme is based on the
waveguide QED system \cite%
{fan1,zhou,fan2,roy,rep,longo,zheng1,zheng2,zheng3,witt,sfan,zhou2,petter,shit,huangjf,yan,brad}%
, in which the strong coupling between the waveguide photons and the
emitters coupled to the waveguide is realized. The routed photons and
control photons are connected by an intermediate single-mode cavity. When
the photons in the coherent state are injected into any of the input ports,
the photon transport does not depend on the phase of the photons. However,
when more than one input ports are injected with coherent-state photons, the
photon transport can be controlled by the phase differences between the
photons injected into different ports. The routed photons and control
photons have the equal mean photon numbers and frequencies. Consequently,
the routed photons can act as the control photons and the control ones can
act as the routed ones. In our router, the mean photon numbers can be either
small or large. Therefore, our router can be realized at the single-photon
level. Under certain conditions, our scheme is a router with two input ports
and two output ports. Compared to \cite{dayang}, the intermediate
single-mode cavity is coupled to both the routed and control photons in our
scheme. This may avoid the cross-contamination of matching different atomic
transitions, respectively, to the routed and control photons.

\begin{figure}[t]
\includegraphics*[width=8cm, height=5cm]{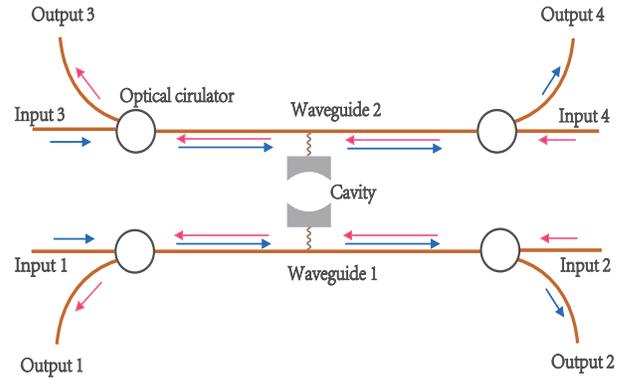}
\caption{Schematic configuration of the all-optical routing of single
photons with four input ports and four output ports. The two optical
waveguides are connected by an optical cavity. Four optical circulators are
employed to separate the input ports from output ports.}
\end{figure}

The system under consideration is depicted in Fig. 1. The cavity is strongly
side-coupled to lossless waveguide $1$ and $2$. The right (left)-moving
photons in waveguide $1$ are connected to the input port $1$ ($2$) and
output port $2$ ($1$) with the optical circulators. And the right
(left)-moving photons in waveguide $2$ are connected to the input port $3$ ($%
4$) and output port $4$ ($3$). The photons injected into any of the input
ports move along the 1D waveguides and then are scattered due to the
photon-cavity interaction. After scattering, the photons may be redirected.
Here we focus on the photon transport influenced by the photon phases. The
system Hamiltonian in the rotating-wave approximation is written as ($\hbar
=1$)%
\begin{eqnarray}
H &=&\sum_{j=1,2}(\int d\omega \omega r_{j\omega }^{\dagger }r_{j\omega
}+\int d\omega \omega l_{j\omega }^{\dagger }l_{j\omega })+\omega
_{c}c^{\dagger }c  \label{hamiltonian} \\
&&+\sum_{j=1,2}\int d\omega g_{j}c^{\dagger }(r_{j\omega }e^{i\omega
z_{c}/v_{g}}+l_{j\omega }e^{-i\omega z_{c}/v_{g}})+h.c.\text{,}  \notag
\end{eqnarray}%
where $r_{j\omega }^{\dagger }$ ($l_{j\omega }^{\dagger })$ creates a
right(left) propagating photon with frequency $\omega $ in the waveguide $j$%
, $c^{\dagger }$ creates a photon in the cavity, $\omega _{c}$ is the cavity
resonance frequency, $g_{j}$ is the coupling strength of the cavity to the
waveguide $j$, $z_{c}$ is the position of the cavity, and $v_{g}$ is the
group velocity of the photons. Here, we have assumed that $g_{j}$ is
frequency-independent, which is equivalent to the Markovian approximation.\
The waveguides are considered with the linear dispersion relation, i.e. $%
\omega =v_{g}\left| k\right| $, with $k$ wave number. We will take $z_{c}$
zero and extend the frequency integration to $\pm \infty $ below.

We study the photon scattering with input-output formalism \cite{cw}. The
input and output operators are defined as $o_{j}^{(in)}(t)=\frac{1}{\sqrt{%
2\pi }}\int d\omega o_{j\omega }(t_{0})e^{-i\omega (t-t_{0})}$ ($o=r,l$) and
$o_{j}^{(out)}(t)=\frac{1}{\sqrt{2\pi }}\int d\omega o_{j\omega
}(t_{1})e^{-i\omega (t-t_{1})}$, respectively. The operator $o_{j\omega
}^{(in)}$ and $o_{j\omega }^{(out)}$ in the scattering theory are related to
the input and output operators through $o_{j}^{(in)}(t)=\frac{1}{\sqrt{2\pi }%
}\int d\omega o_{j\omega }^{(in)}e^{-i\omega t}$ and $o_{j}^{(in)}(t)=\frac{1%
}{\sqrt{2\pi }}\int d\omega o_{j\omega }^{(in)}e^{-i\omega t}$ \cite{sfan},
respectively. The system initial state $\left| \Psi _{0}\right\rangle $ is a
simple product state of the two waveguide field states and the cavity state.
In our scheme, initially, the cavity is in the vacuum state and the injected
photons are in the coherent states. For the coherent input state, $%
o_{j}^{(in)}(t)\left| \Psi _{0}\right\rangle =\frac{1}{\sqrt{2\pi }}\int
d\omega \alpha _{\omega }e^{-i\omega t}\left| \Psi _{0}\right\rangle $, with
$\alpha _{\omega }$ being a complex number. The mean number of the
coherent-state photons is represented by $\int d\omega \left| \alpha
_{\omega }\right| ^{2}$. By the input-output formalism, we find%
\begin{eqnarray}
o_{j}^{(out)}(t) &=&o_{j}^{(in)}(t)-i\sqrt{\gamma _{j}}c(t)\text{,}
\label{motion} \\
\dot{c}(t) &=&(-i\omega _{c}-\sum_{j}\gamma _{j})c(t)-i\sum_{j,o}\sqrt{%
\gamma _{j}}o_{j}^{(in)}(t)\text{,}  \notag
\end{eqnarray}%
with $\gamma _{j}=2\pi g_{j}^{2}$ being the decay rates from the cavity to
the waveguide $j$. From Eqs. (2), both the expectation values $\left\langle
\Psi _{0}\right| o_{j}^{(out)\dagger }(t)o_{j}^{(out)}(t)\left| \Psi
_{0}\right\rangle $ and $\left\langle \Psi _{0}\right| o_{j\omega
}^{(out)\dagger }o_{j\omega }^{(out)}\left| \Psi _{0}\right\rangle $ can be
obtained under the initial conditions.

We first consider the case that the photons with frequency $\omega $ in a
coherent state with mean photon number $\left| \alpha \right| ^{2}$ are
injected into input port $1$. After calculations, we obtain $%
o_{j}^{(out)}(t)\left| \Psi _{0}\right\rangle =f_{oj}(\gamma _{1},\gamma
_{1},\delta _{k},\omega _{c},\omega )e^{-i\omega t}\left| \Psi
_{0}\right\rangle $. Therefore, the output photons have the same frequency
with the input photons due to the conversation of energy. The mean numbers
of the photons outputting from each ports are%
\begin{eqnarray}
N_{r1}^{(out)} &=&\frac{\delta _{\omega }^{2}+\gamma _{2}^{2}}{\delta
_{\omega }^{2}+(\gamma _{1}+\gamma _{2})^{2}}\left| \alpha \right| ^{2}\text{%
,}  \label{singleinput} \\
N_{l1}^{(out)} &=&\frac{\gamma _{1}^{2}}{\delta _{\omega }^{2}+(\gamma
_{1}+\gamma _{2})^{2}}\left| \alpha \right| ^{2}\text{,}  \notag \\
N_{r2}^{(out)} &=&N_{l2}^{(out)}=\frac{\gamma _{1}\gamma _{2}}{\delta
_{\omega }^{2}+(\gamma _{1}+\gamma _{2})^{2}}\left| \alpha \right| ^{2}\text{%
,}  \notag
\end{eqnarray}%
with $\delta _{\omega }=\omega _{c}-\omega $ the detuning. $N_{rj}^{(out)}$ (%
$N_{lj}^{(out)}$) is the mean number of the right (left)-moving photons in
the waveguide $j$ after scattering. Hence, $N_{r1}^{(out)}$, $N_{l1}^{(out)}$%
, $N_{r2}^{(out)}$ and $N_{l2}^{(out)}$ correspond to the mean numbers of
the photons outputting from ports $2$, $1$, $4$ and $3$, respectively. It is
easy to verify the conservation relation $\sum\limits_{o,j}N_{oj}^{(out)}=%
\left\langle r_{1}^{(in)\dagger }(t)r_{1}^{(in)}(t)\right\rangle =\left|
\alpha \right| ^{2}$. When the input photons resonantly interact with the
cavity and the coupling strengths of the cavity to the two waveguides are
equal, i.e. $\delta _{\omega }=0$ and $\gamma _{1}=\gamma _{2}$, the photons
are redirected into the four output ports equally. When $\delta _{\omega
}\gg \gamma _{1}$ or $\gamma _{2}\gg \gamma _{1}$, the waveguide $1$ is
almost decoupled to the cavity and we find $N_{r1}^{(out)}\rightarrow \left|
\alpha \right| ^{2}$. When the cavity is decoupled to the waveguide $2$ and
the input photons resonantly interact with the cavity, the photons are
completely reflected and redirected into the output port $1$.

The photons injected into different ports arrive at the position $z_{c}$\
simultaneously and then interact with the intermediate cavity.\ We proceed
to study the routing of the photons by photons in two cases. One case is the
routing of photons by photons injected into another input ports, the other
case is by photons injected into other two input ports.

\begin{figure}[t]
\includegraphics*[width=6cm, height=4cm]{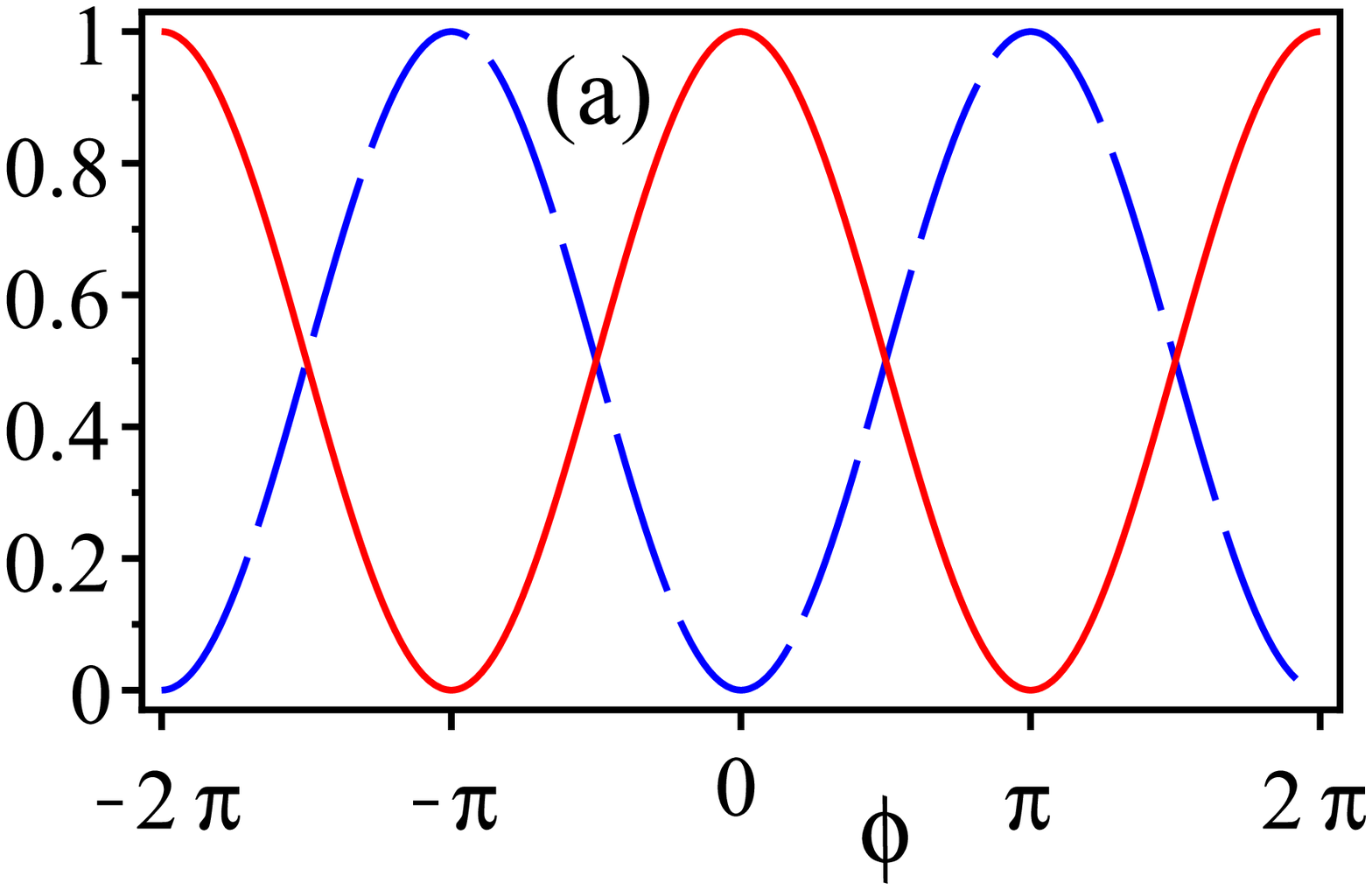} \includegraphics*%
[width=6cm, height=4cm]{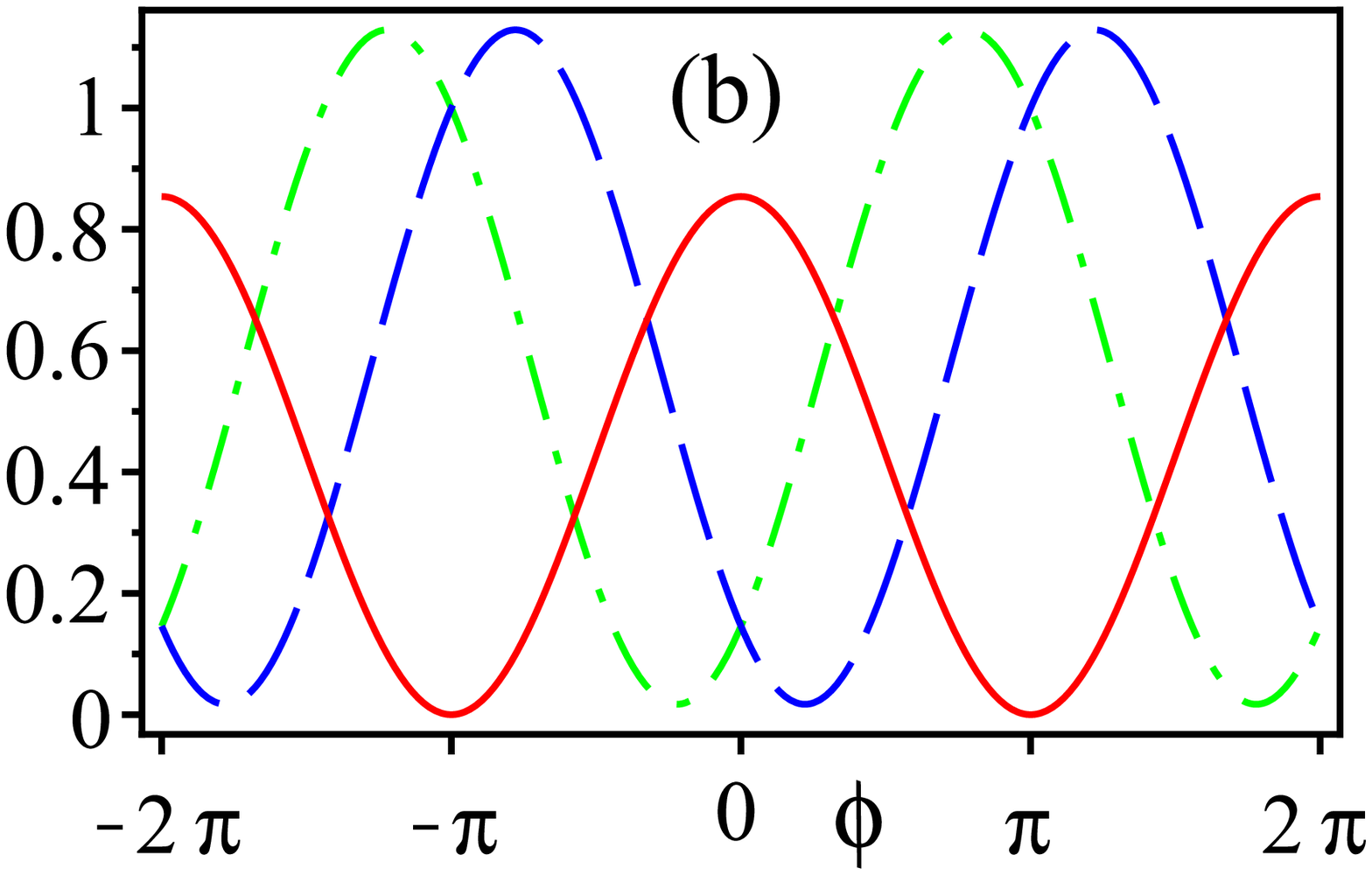} \includegraphics*[width=6cm, height=4cm]{%
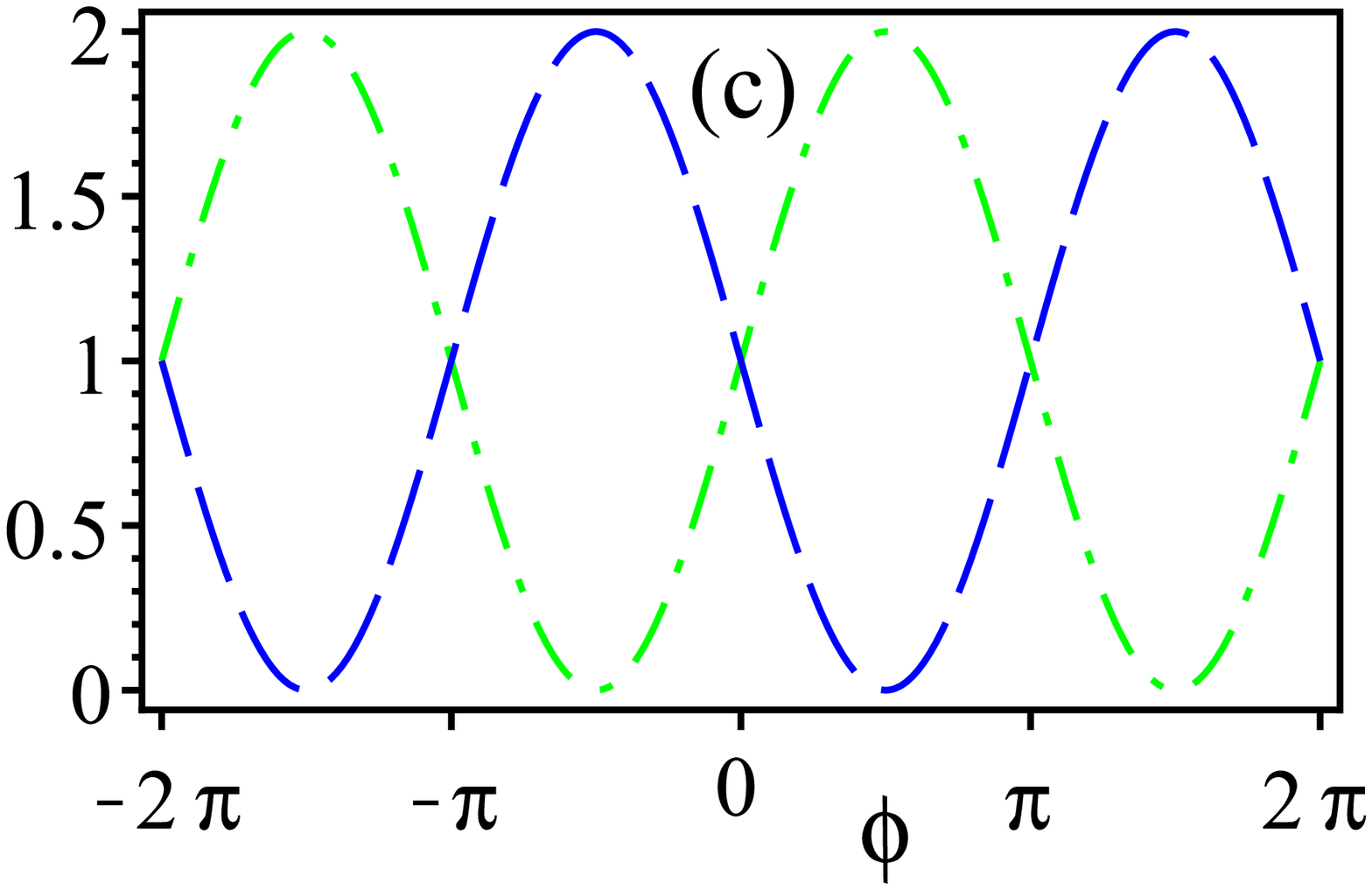}
\caption{The mean photon numbers $N_{oj}^{(out)}$ against the phase
differences $\protect\theta $ in the two-input case. The blue dashed lines
are $N_{r1}^{(out)}$, the green dashed dotted lines are $N_{l1}^{(out)}$,
the red dotted lines are $N_{r2}^{(out)}=N_{l2}^{(out)}$. We take $\protect%
\delta _{\protect\omega }=0$, $\protect\gamma _{2}=\protect\gamma _{1}$ in
(a), $\protect\delta _{\protect\omega }=0.5\protect\gamma _{1}$, $\protect%
\gamma _{2}=0.6\protect\gamma _{1}$, in (b), and $\protect\delta _{\protect%
\omega }=\protect\gamma _{1}$, $\protect\gamma _{2}=0$ in (c). The other
parameter is $\left| \protect\alpha \right| ^{2}=1$. All the parameters but $%
\left| \protect\alpha \right| ^{2}$ are in units of $\protect\gamma _{1}$.
The mean photon number $N_{l1}^{(out)}$ is not given in (a) because $%
N_{r1}^{(out)}=N_{l1}^{(out)}$.}
\end{figure}

\textit{Two-input case.}---In the two-input case, it is enough to study the
situations that photons are injected into port $1$ and $2$ and that the
photons are injected into port $1$ and $3$. This can be understood by the
expression of Hamiltonian (1). When the photons in the coherent states are
injected into port $1$ and $2$, the mean numbers of the output photons are
obtained as%
\begin{eqnarray}
N_{r1}^{(out)} &=&[1-2\frac{(1+\cos \phi )\gamma _{1}\gamma _{2}+\gamma
_{1}\delta _{\omega }\sin \phi }{\delta _{\omega }^{2}+(\gamma _{1}+\gamma
_{2})^{2}}]\left| \alpha \right| ^{2}\text{,}  \label{twoinput} \\
N_{l1}^{(out)} &=&[1-2\frac{(1+\cos \phi )\gamma _{1}\gamma _{2}-\gamma
_{1}\delta _{\omega }\sin \phi }{\delta _{\omega }^{2}+(\gamma _{1}+\gamma
_{2})^{2}}]\left| \alpha \right| ^{2}\text{,}  \notag \\
N_{r2}^{(out)} &=&N_{l2}^{(out)}=\frac{2(1+\cos \phi )\gamma _{2}\gamma _{1}%
}{\delta _{\omega }^{2}+(\gamma _{1}+\gamma _{2})^{2}}\left| \alpha \right|
^{2}\text{,}  \notag
\end{eqnarray}%
with $\phi $ being the phase difference between the photons injected into
different ports. Here we have taken that the photons injected into the two
input ports have the same photon mean number $\left| \alpha \right| ^{2}$
and the same frequency $\omega $. Similar to the single-input case, the
output photons have the same frequency with the input photons. It is
interesting that the expressions of the mean output-photon numbers are
periodic functions of $\phi $ with period $2\pi $. Therefore, the routing of
photons can be achieved by the phase of other photons injected into another
input port. When $\phi =2\pi $, $\delta _{\omega }=0$ and $\gamma
_{1}=\gamma _{2}$, the photons are completely redirected into output ports $%
3 $ and $4$ due to the constructive interference. However, when $\phi =\pi $%
, the photons are completely redirected into output ports $1$ and $2$ due to
the destructive interference. To see the details of the routing property, we
plot the mean photon numbers in Eqs. (3) against the phase difference\ in
fig. 2. Therefore, the routing of the coherent-state photons injected into
the input port $1$ can be achieved by the phase of the coherent-state
photons injected into the input port $2$. In our scheme, this routing is
based on the interferences determined by the phase difference. These
interferences can not be obtained when the input photons are in Fock states %
\cite{petter}. This is because the coherent state is the eigenstate of the
annihilation operator.

When the cavity is decoupled to the waveguide $2$, i.e. $\gamma _{2}=0$, our
scheme becomes a router with two input and two output ports. The mean
numbers of the photons outputting from either port are obtained as $%
N_{r1}^{(out)}=\frac{\delta _{\omega }^{2}+\gamma _{1}^{2}-2\gamma _{1}\sin
\phi \delta _{\omega }}{\delta _{\omega }^{2}+\gamma _{1}^{2}}\left| \alpha
\right| ^{2}$, and $N_{l1}^{(out)}=\frac{\delta _{\omega }^{2}+\gamma
_{1}^{2}+2\gamma _{1}\sin \phi \delta _{\omega }}{\delta _{\omega
}^{2}+\gamma _{1}^{2}}\left| \alpha \right| ^{2}$. It is interesting that
when $\delta _{\omega }^{2}=\gamma _{1}^{2}$, the expectation value $%
N_{o1}^{(out)}$ can be from $0$ to $2\left| \alpha \right| ^{2}$ by
adjusting the phase $\phi $. The details are shown in Fig. 2c.

When the photons in coherent states are injected into input ports $1$ and $3$%
, the outcomes have the forms similar to the outcomes in Eqs. (4) except $%
\gamma _{j}$. Hence, it is not necessary to study the details of this
situation. When $\gamma _{1}=\gamma _{2}$, the outcomes are equal to the
outcomes in Eqs. (4). Consequently, under the conditions $\gamma _{1}=\gamma
_{2}$ and $\delta _{\omega }=0$, the photons can be completely directed into
output ports $2$ ($1$) and $4$ ($2$) when $\phi =\pi $ ($\phi =2\pi $).

\begin{figure}[t]
\includegraphics*[bb=3 119 574 433, width=4.2cm, height=2.5cm]{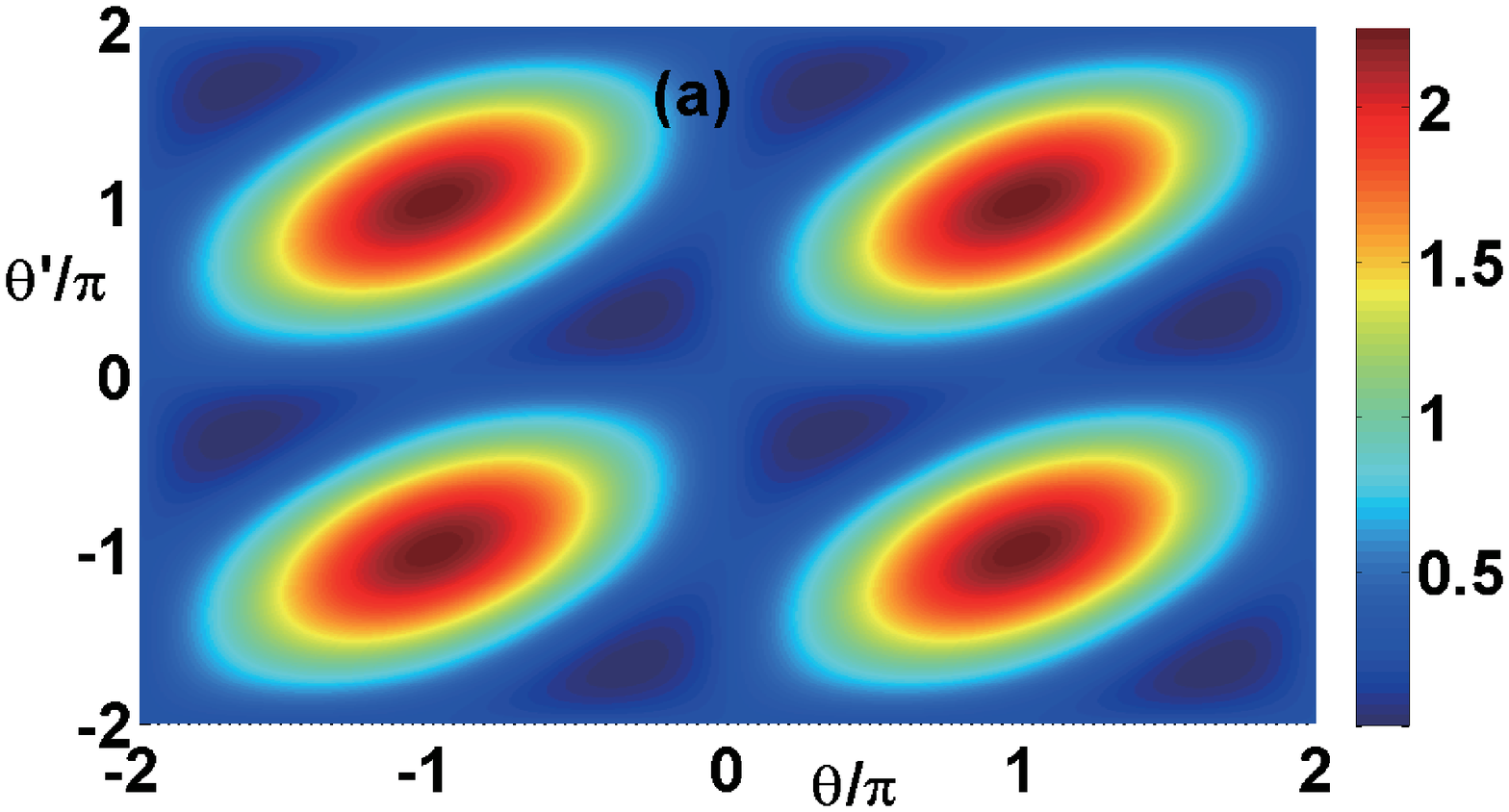} \includegraphics*%
[bb=3 119 574 433,width=4.2cm, height=2.5cm]{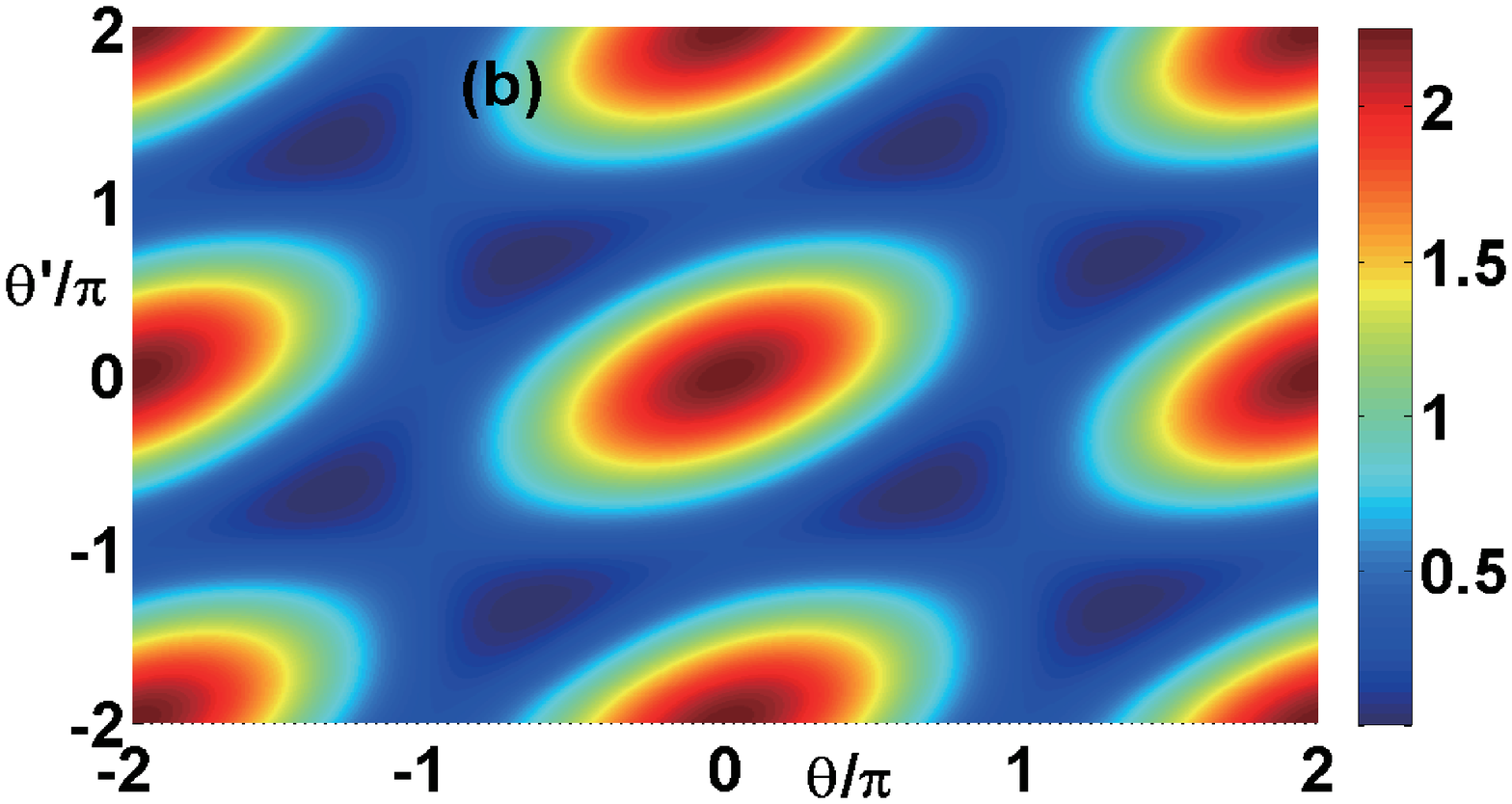} \includegraphics*[bb=3 119 574 433,width=4.2cm,
height=2.5cm]{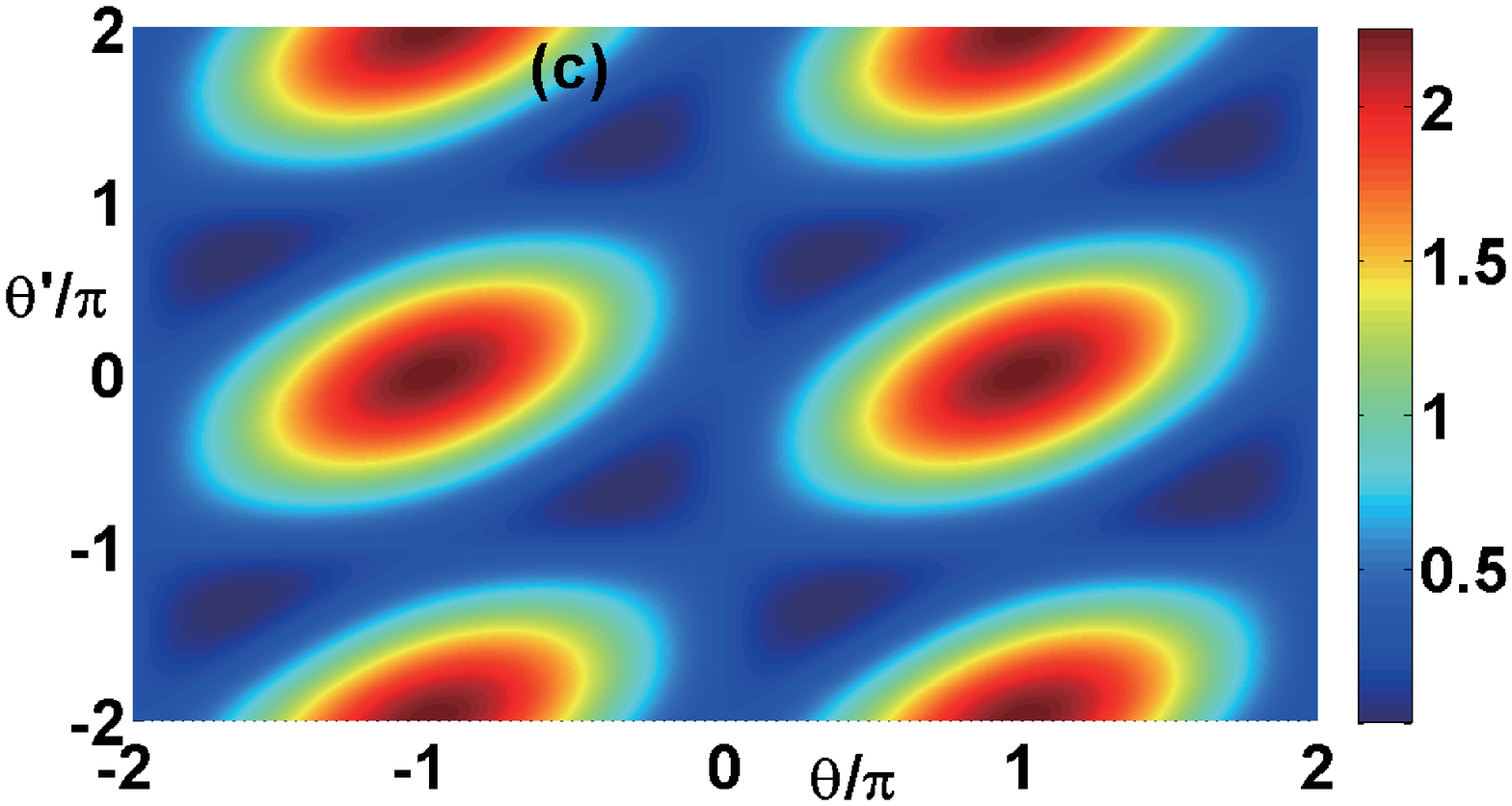} \includegraphics*[bb=3 119 574 433,width=4.2cm, height=2.5cm]{%
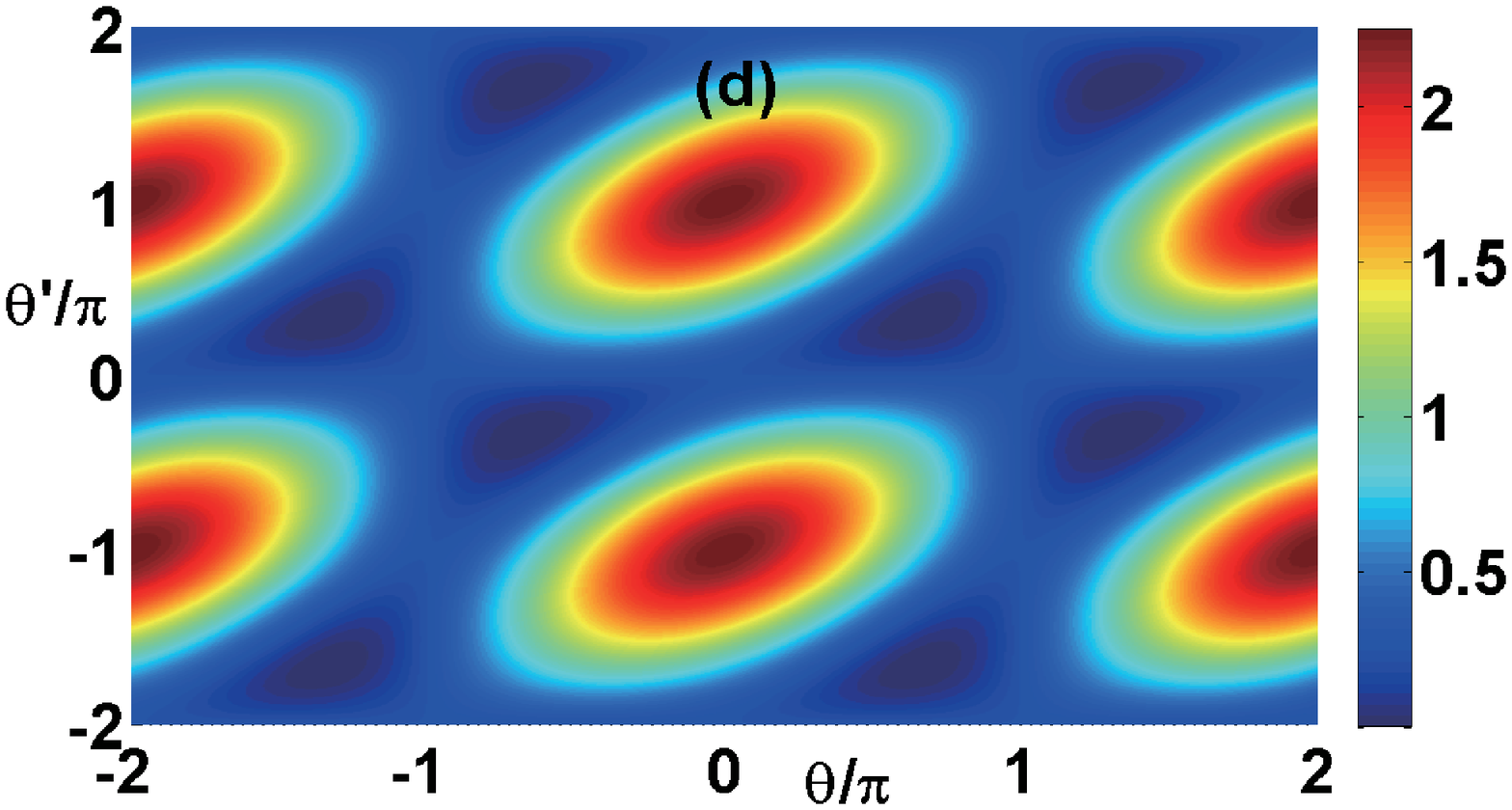}
\caption{The mean photon numbers against the phase differences in the
three-input case. (a), (b), (c) and (d) denote $N_{rl}^{(out)}$, $%
N_{l1}^{(out)}$, $N_{r2}^{(out)}$ and $N_{l2}^{(out)}$, respectively. For
all the plots, the parameters are $\protect\delta _{k}=0$, $\protect\gamma %
_{2}=\protect\gamma _{1}$, and $\left| \protect\alpha \right| ^{2}=1$. All
the parameters but $\left| \protect\alpha \right| ^{2}$ are in units of $%
\protect\gamma _{1}$.}
\end{figure}

\textit{Three-input case.}---In the three-input case, it is enough to study
the situation that the photons are injected into input ports $1$, $3$ and $4$%
. When the coherent-state photons with frequency $\omega $ are injected into
the input ports $1$, $3$ and $4$, the output photons have the frequency $%
\omega $ and the mean numbers of the output photons are obtained as

\begin{eqnarray}
N_{r1}^{(out)} &=&\frac{%
\begin{array}{c}
\delta _{\omega }^{2}+\gamma _{2}^{2}+2\gamma _{1}\gamma _{2}-2\delta
_{\omega }\sqrt{\gamma _{1}\gamma _{2}}(\sin \theta \\
+\sin \theta ^{\prime })-2\sqrt{\gamma _{1}\gamma _{2}}\gamma _{2}(\cos
\theta +\cos \theta ^{\prime }) \\
+2\cos (\theta -\theta ^{\prime })\gamma _{1}\gamma _{2}%
\end{array}%
}{\delta _{\omega }^{2}+(\gamma _{1}+\gamma _{2})^{2}}\left| \alpha \right|
^{2}\text{,}  \label{threeinput} \\
N_{l1}^{(out)} &=&\frac{%
\begin{array}{c}
\gamma _{1}^{2}+2[1+\cos (\theta -\theta ^{\prime })]\gamma _{1}\gamma _{2}
\\
+2(\cos \theta +\cos \theta ^{\prime })\gamma _{1}\sqrt{\gamma _{1}\gamma
_{2}}%
\end{array}%
}{\delta _{\omega }^{2}+(\gamma _{1}+\gamma _{2})^{2}}\left| \alpha \right|
^{2}\text{,}  \notag \\
N_{r2}^{(out)} &=&\frac{%
\begin{array}{c}
\delta _{\omega }^{2}+(\gamma _{1}+\gamma _{2})^{2}-\gamma _{2}\gamma
_{1}+2\sin \theta \sqrt{\gamma _{1}\gamma _{2}}\delta _{\omega } \\
+2\sin (\theta -\theta ^{\prime })\gamma _{2}\delta _{\omega }-2\cos \theta
\gamma _{1}\sqrt{\gamma _{1}\gamma _{2}} \\
+2\cos \theta ^{\prime }\gamma _{2}\sqrt{\gamma _{1}\gamma _{2}}-2\cos
(\theta ^{\prime }-\theta )\gamma _{1}\gamma _{2}%
\end{array}%
}{\delta _{\omega }^{2}+(\gamma _{1}+\gamma _{2})^{2}}\left| \alpha \right|
^{2}\text{,}  \notag \\
N_{l2}^{(out)} &=&\frac{%
\begin{array}{c}
\delta _{\omega }^{2}+(\gamma _{1}+\gamma _{2})^{2}-\gamma _{2}\gamma
_{1}+2\delta _{\omega }[\sin \theta ^{\prime }\sqrt{\gamma _{1}\gamma _{2}}
\\
+\sin (\theta ^{\prime }-\theta )\gamma _{2}]-2\cos (\theta -\theta ^{\prime
})\gamma _{2}\gamma _{1} \\
-2\cos \theta ^{\prime }\gamma _{1}\sqrt{\gamma _{1}\gamma _{2}}+2\cos
\theta \gamma _{2}\sqrt{\gamma _{1}\gamma _{2}}%
\end{array}%
}{\delta _{\omega }^{2}+(\gamma _{1}+\gamma _{2})^{2}}\left| \alpha \right|
^{2}\text{,}  \notag
\end{eqnarray}%
with $\theta $ ($\theta ^{\prime }$) the phase difference between the
photons injected into input ports $1$ and $3$ ($4$).The photons injected
into each of the three ports have the same mean photon number $\left| \alpha
\right| ^{2}$. The mean numbers of output photons in Eqs. (5) against the
phase differences are plotted in Fig. 3. It shows that the routing of the
photons by other photons can be achieved in the three-input case.

We note that although we have taken the mean photon number $|\alpha |^{2}=1$
in all the plots, the routing properties do not depend on $|\alpha |^{2}$.
This can be understood from the expressions of Eqs. (4) and (5). Hence, the
routing can be achieved at the single-photon level.

We have studied the routing of photons when the input photons are in
single-mode coherent states, without considering the cavity decay to other
modes but the waveguide modes. The cavity decay can be incorporated by
introducing the nonhermitian Hamiltonian $H_{non}=-i\gamma _{c}c^{\dagger }c$%
, with $\gamma _{c}$ the decay rate. The injected coherent-state prepared in
a Gaussian-type wave packet is defined as $a_{\omega }^{(in)}\left| \Psi
_{0}\right\rangle =\alpha _{\omega }\left| \Psi _{0}\right\rangle $. The
complex number $\alpha _{k}$ has the form of $\alpha _{\omega }=\frac{\alpha
}{\sqrt[4]{2\pi \Omega ^{2}}}e^{-\frac{(\omega -\omega _{0})^{2}}{4\Omega
^{2}}}$, with $2\Omega $ the bandwidth and $\omega _{0}$ the center
frequency. The mean photon number of the wave packet is $\int d\omega \left|
\alpha _{\omega }\right| ^{2}=\left| \alpha \right| ^{2}$. For the
Gaussian-type wave-packet input, the mean output-photon numbers can be
obtained by numerical evaluations. We plot the routing property when the
photons in the coherent state prepared in Gaussian-type wave packet are
injected into input ports $1$ and $2$ in Fig. 4. In Fig. 4, the cavity decay
has been incorporated. In Fig. 4(a), the up bound of $%
N_{r1}^{(out)}=N_{l1}^{(out)}$ is barely affected but the up bound of $%
N_{r2}^{(out)}=N_{l2}^{(out)}$ decreases evidently compared to Fig. 2. In
Fig. 4(c), both the up bound $N_{r1}^{(out)}$ and $N_{l1}^{(out)}$\ decrease
evidently. These are mainly due to the fact that we have considered the
wave-packet bandwidth, which can be understood as follows. The
frequency-dependent condition $\delta _{\omega }=0$ is necessary when the
value of $N_{r2}^{(out)}$ in Fig. 4(a) reaches unit. However, the unit value
of $N_{r1}^{(out)}$ in 4(a) only needs the condition $\theta =\pi $, which
is frequency-independent. In 4(c), we take $\delta _{\omega }=\gamma _{1}$,
which is frequency-dependent. The outcomes obtained under the
frequency-dependent condition are affected by the bandwidth. The effect
caused by the cavity decay can be studied by the mean number $N^{(out)}$ of
all the output photons, with $%
N^{(out)}=N_{r1}^{(out)}+N_{l1}^{(out)}+N_{r2}^{(out)}+N_{l2}^{(out)}$. As
is shown in Fig. 4(d), when $\theta =\pi $, the $N^{(out)}$ is not affected
by the cavity decay due to the destructive interference. However, when $%
\theta =2\pi $, the cavity decay has obvious effect due to the constructive
interference.
\begin{figure}[h]
\includegraphics*[width=4.2cm, height=2.5cm]{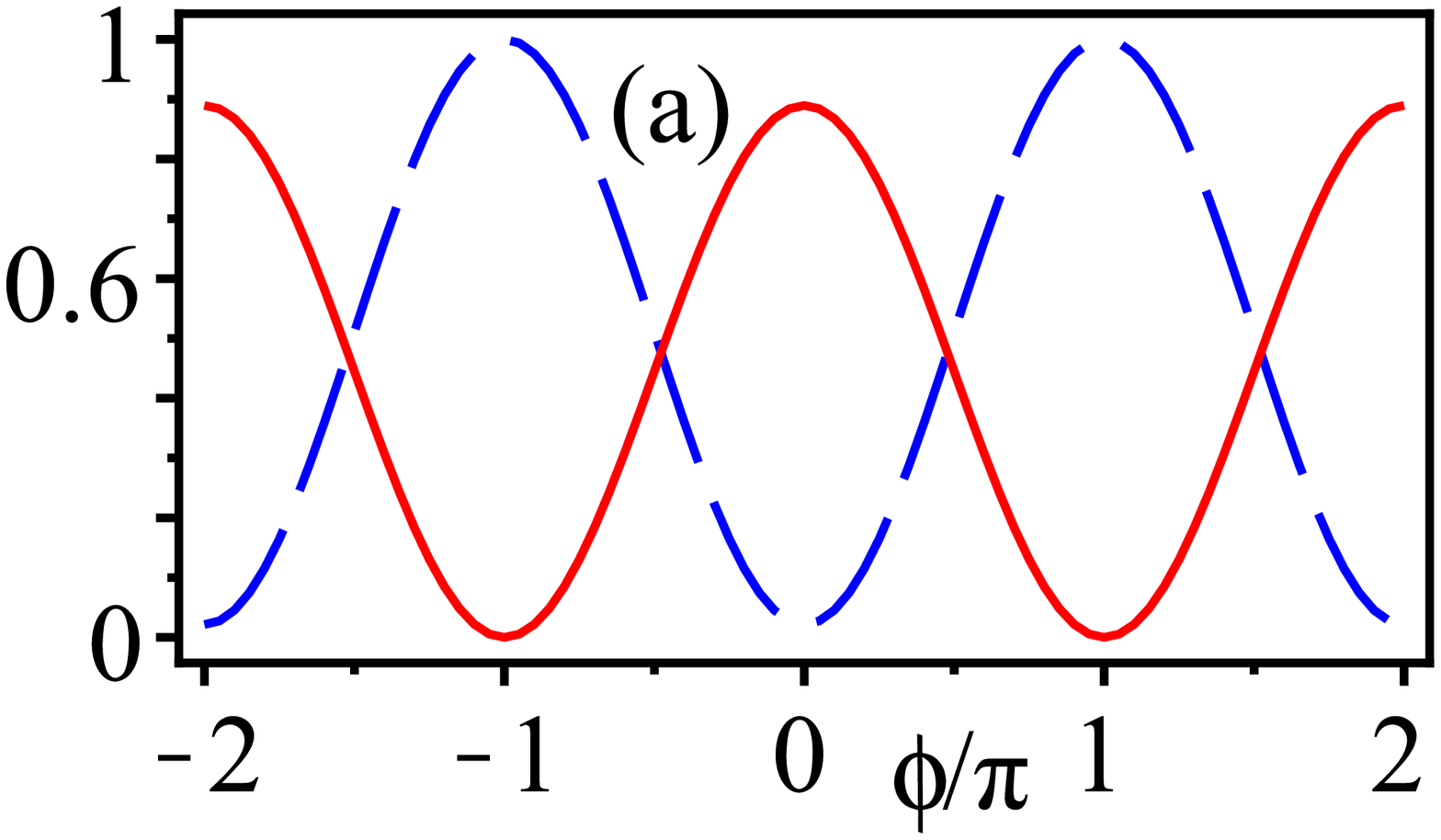} \includegraphics*%
[width=4.2cm, height=2.5cm]{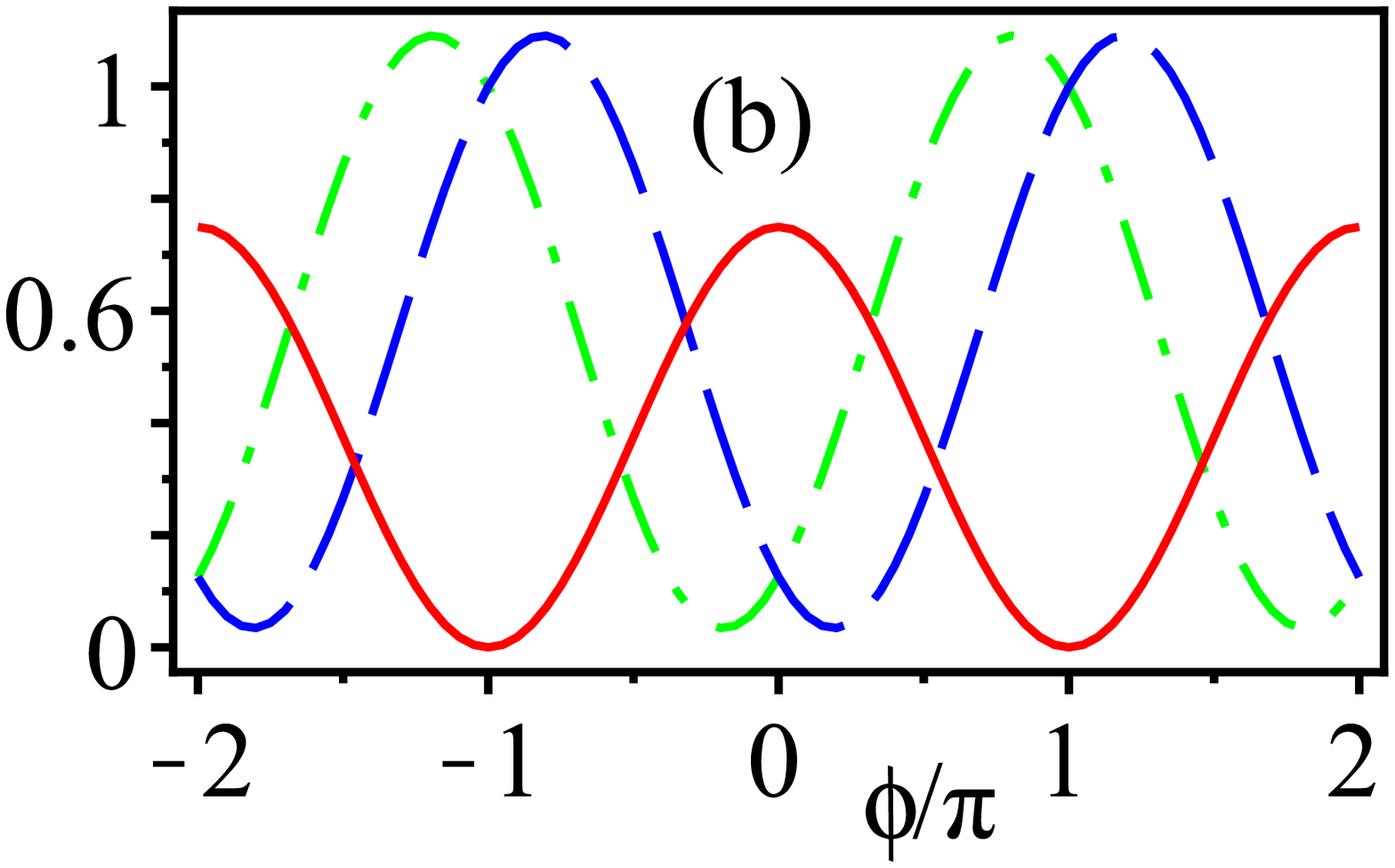} \includegraphics*[width=4.2cm,
height=2.5cm]{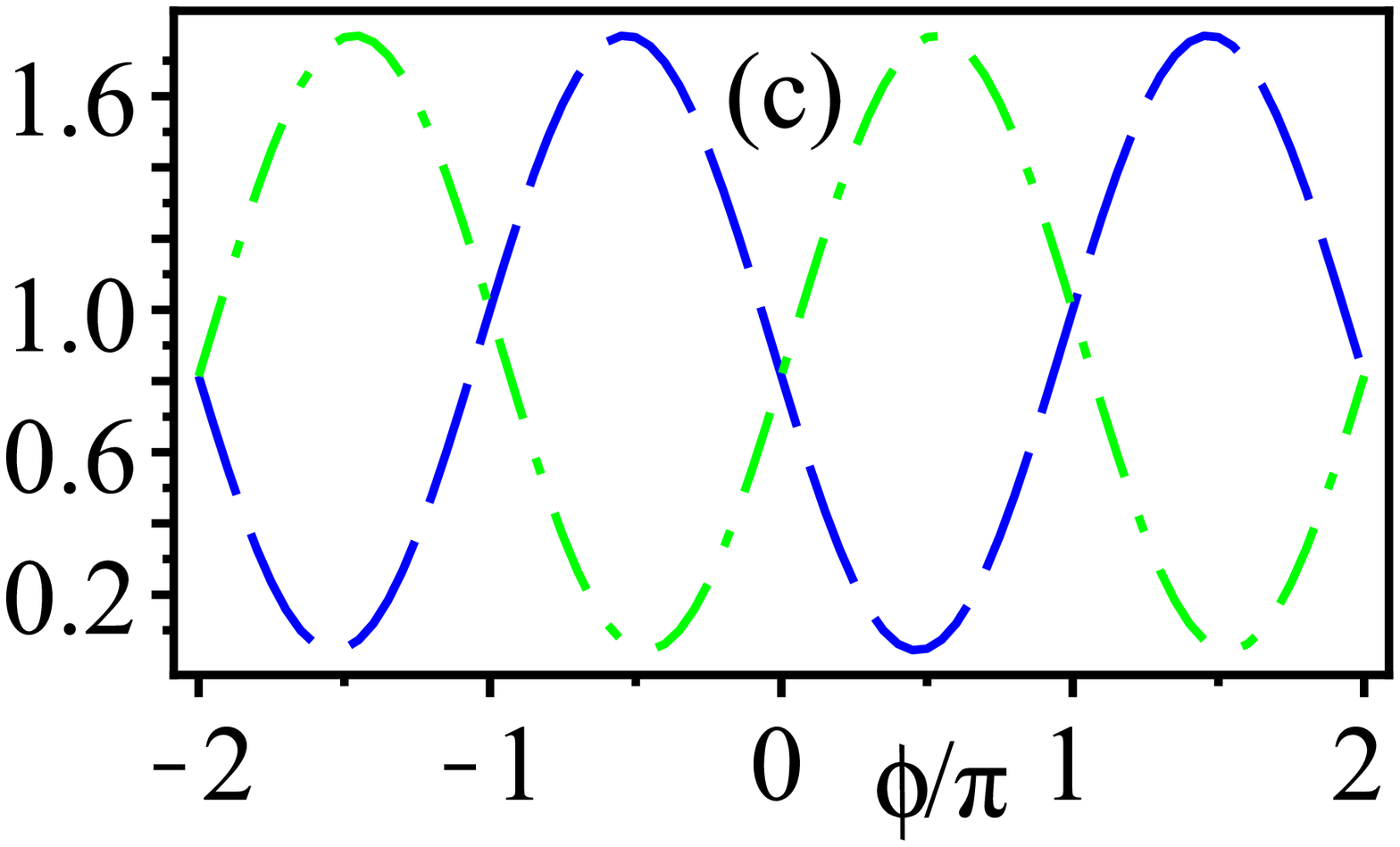} \includegraphics*[width=4.2cm, height=2.5cm]{%
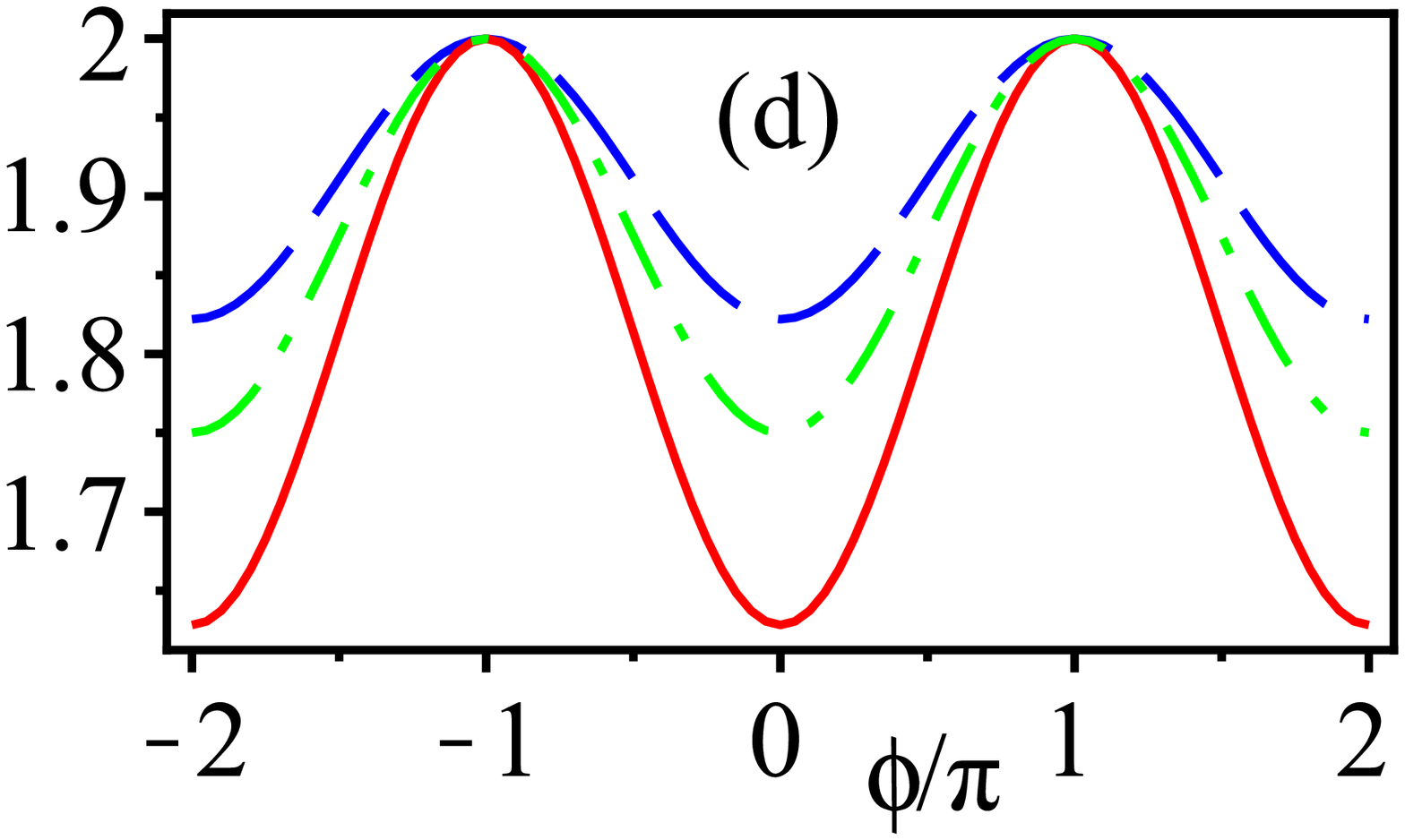}
\caption{The mean numbers of the output photons against the phase
differences in the two-input case. The input photons are in coherent states
prepared in Gaussian-type wavepackets and the cavity decay has been
considered. For all the plots, we take $\Omega =0.3\protect\gamma _{1}$ and $%
\protect\gamma _{c}=0.1\protect\gamma _{1}$.\ In (a), (b) and (c), the blue
dashed lines are $N_{r1}^{(out)}$, the green dashed dotted lines are $%
N_{l1}^{(out)}$, the red dotted lines are $N_{r2}^{(out)}=N_{l2}^{(out)}$.
The other parameters in (a), (b) and (c) are the same to the parameters in
Fig. 2(a), 2(b) and 2(c), respectively. (d) shows the sum of the mean
numbers of the photons outputting from all output ports $N^{(out)}$. The
blue dashed line, green dashed dotted line, and red solid lines in (d)
correspond to the situations as shown in (a), (b) and (c), respectively.}
\end{figure}

In conclusion, we have presented a detailed investigation on the routing of
single photons with four input ports and four output ports by single
photons. The routing is achieved by the interferences related to the phase
differences between the coherent-state photons. The routed photons can play
the role of control photons, and the control photons can also play the role
of routed photons. Our scheme is of significance to build the quantum
networks. We hope that this routing will be achieved experimentally in the
near future.

This work is supported by '973'\ program (2010CB922904), grants from Chinese
Academy of Sciences, NSFC No. 11175248 and No. 11474044.

\end{document}